\documentclass[aps,
prb,
reprint,
amsmath,amssymb,
superscriptaddress,
showpacs]{revtex4-1}
\usepackage{CJK}
\usepackage{bm}
\usepackage{epsfig}
\usepackage[caption=false]{subfig}
\begin{document}

%\preprint{}
\begin{CJK*}{UTF8}{gbsn} % chinese characters

\title{Magnetism of monomer MnO and heterodimer FePt@MnO nanoparticles}

\author{X.~Sun (孙笑)}
\affiliation{J\"ulich Centre for Neutron Science JCNS and Peter Gr\"unberg Institut PGI, JARA-FIT, Forschungszentrum J\"ulich GmbH, 52425 J\"ulich, Germany
}
\author{A.~Klapper}
\affiliation{J\"ulich Centre for Neutron Science JCNS and Peter Gr\"unberg Institut PGI, JARA-FIT, Forschungszentrum J\"ulich GmbH, 52425 J\"ulich, Germany 
}
\author{Y.~Su (苏夷希)}
\affiliation{J\"ulich Centre for Neutron Science JCNS at Heinz Maier-Leibnitz Zentrum MLZ, Forschungszentrum J\"ulich GmbH, 85747 Garching, Germany
}
\author{K.~Nemkovski}
\affiliation{J\"ulich Centre for Neutron Science JCNS at Heinz Maier-Leibnitz Zentrum MLZ, Forschungszentrum J\"ulich GmbH, 85747 Garching, Germany
}
\author{A.~Wildes}
\affiliation{Institut Laue-Langevin, BP 156, 38042 Grenoble Cedex 9, France
}
\author{H.~Bauer}
\affiliation{Institut f\"ur Anorganische Chemie und Analytische Chemie, Johannes Gutenberg-Universit\"at Mainz, 55099 Mainz, Germany
}
\author{O.~K\"ohler}
\affiliation{Institut f\"ur Anorganische Chemie und Analytische Chemie, Johannes Gutenberg-Universit\"at Mainz, 55099 Mainz, Germany
}
\author{A.~Schilmann}
\affiliation{Institut f\"ur Anorganische Chemie und Analytische Chemie, Johannes Gutenberg-Universit\"at Mainz, 55099 Mainz, Germany
}
\author{W.~Tremel}
\affiliation{Institut f\"ur Anorganische Chemie und Analytische Chemie, Johannes Gutenberg-Universit\"at Mainz, 55099 Mainz, Germany
}
\author{O.~Petracic}
\affiliation{J\"ulich Centre for Neutron Science JCNS and Peter Gr\"unberg Institut PGI, JARA-FIT, Forschungszentrum J\"ulich GmbH, 52425 J\"ulich, Germany 
}
\author{Th.~Br\"uckel}
\affiliation{J\"ulich Centre for Neutron Science JCNS and Peter Gr\"unberg Institut PGI, JARA-FIT, Forschungszentrum J\"ulich GmbH, 52425 J\"ulich, Germany 
}

\date{\today}

\begin{abstract}
We report about the magnetic properties of antiferromagnetic (AF) MnO nanoparticles (NPs) with different sizes (6-19~nm). Using a combination of polarized neutron scattering and magnetometry we were able to resolve previously observed peculiarities. Magnetometry, on the one hand, reveals a peak in the zero field cooled (ZFC) magnetization curves at low temperatures ($\sim$25~K) but \textit{no} feature around the N\'eel temperature at 118~K. On the other hand, polarized neutron scattering shows the expected behavior of the AF order parameter vanishing around 118~K. Moreover, hysteresis curves measured at various temperatures reveal an exchange bias effect indicating a coupling of an AF core to a ferromagnetic (FM)-like shell. ZFC data measured at various fields exclude a purely superparamagnetic (SPM) scenario. We conclude that the magnetic behavior of MnO particles can be explained by a superposition of SPM-like thermal fluctuations of the AF-N\'eel vector inside the AF core \textit{and} a strong magnetic coupling to a ferrimagnetic Mn$_2$O$_3$ or Mn$_3$O$_4$ shell. In addition, we have studied heterodimer (`Janus') particles, where a FM FePt particle is attached to the AF MnO particle. Via the exchange bias effect, the magnetic moment of the FePt subunit is stabilized by the MnO.
\end{abstract}


\maketitle
\end{CJK*}

\section{Introduction}\label{Introduction}
Magnetic NPs have attracted strong interest for decades due to their potential applications in magnetic data storage, ferrofluidic systems and nanomedicine \cite{Sun2006,Zahn2001,Bedanta2013,Terris2005}. In fundamental research in particular spherical magnetic NPs constitute model systems to study finite size \cite{Batlle2002} and spin canting effects \cite{Disch2012}. Due to various surface effects, spin-glass-like behavior \cite{Tiwari2005,Salabas2006,Winkler2005,Yi2007}, core-shell interaction \cite{Salabas2006,Nogues2005,Winkler2008} or weak ferromagnetism \cite{Tomou2006} has been found. Moreover, with control of the size and the shape of the NPs, the structure of 2D and 3D periodic assemblies of NPs can be manipulated \cite{Disch2011,Disch2013,Wetterskog2016a} thus opening a way to fabricate novel materials with specific properties. In particular, a heterodimer NP composed of two different NPs in close contact is a potential candidate to achieve \textit{multifunctional} materials combining various physical properties, e.g. magnetic, electronic, and optical properties.

MnO is AF with a bulk N\'eel temperature of $T_N$~=~118~K and rocksalt crystal structure at room temperature \cite{Shull1951}. X-ray diffraction, neutron scattering as well as magnetometry experiments have been performed on MnO NPs with various sizes \cite{Chatterji2010,Feygenson2010,Wang2011,Golosovsky2001}. The AF order of MnO NPs has recently been studied using neutron scattering \cite{Chatterji2010,Feygenson2010,Wang2011,Golosovsky2001}. A rounding of the magnetic phase transition in contrast to the first-order transition of bulk MnO has been observed \cite{Feygenson2010,Wang2011,Golosovsky2001}.

Despite numerous studies on MnO bulk and MnO NPs, this system is not well understood. E.g. a peculiar peak at low temperatures in the ZFC curve is often found in magnetometry experiments of MnO NPs \cite{Chatterji2010}. Usually it is attributed to SPM behavior \cite{Bedanta2013,Batlle2002,Wiekhorst2003,Dormann1997,Bean1959}, which seems doubtful because SPM is based on the thermally excited switching of a ferro- or ferrimagnetically ordered monodomain \cite{Bedanta2013,Batlle2002,Wiekhorst2003,Dormann1997,Bean1959}. However, neutron scattering results have shown AF order below the N\'eel temperature \cite{Feygenson2010}. In addition, it is well known that a peak in the ZFC curve and a splitting of the ZFC and field cooled (FC) curves can also arise from several other types of systems, e.g. spin glasses\cite{Winkler2008,Mydosh1993}, superspin glasses \cite{Petracic2006,Petracic2010}, diluted antiferromagnets \cite{Benitez2011a} and even ferromagnets \cite{Ebbing2011}. We will show in this paper that this peak can in fact not be attributed to pure SPM behavior, but rather emerges from antiferro-superparamagnetic (AF-SPM) behavior combined with an AF-FM core-shell structure of the MnO particles.\\
\indent In addition, heterodimer NPs composed of an AF MnO NP in close contact with a FM FePt NP are a novel type of a multifunctional nanomagnet \cite{Schladt2012}. At the interface between the FePt NPs and the MnO NPs, an exchange bias effect \cite{Nogues1999,Berkowitz1999} occurs. The FM spins in FePt NPs are magnetically stabilized by the MnO NPs subunits and thus the blocking temperature of the FePt NPs is increased \cite{Schladt2012}.
\section{Experimental}\label{Experimental}
\subsection{Synthesis of monomer MnO NPs}\label{syn_MnO}
MnO NPs were synthesized by thermal decomposition of a manganese oleate precursor according to the procedure described in Ref.~\cite{Schladt2009} with few modifications: 1.24~g of manganese oxide precursor were dissolved in 10~mL of 1-octadecene and degassed three times at 80$^{\circ}$C under reduced pressure (1$\times$10$^{-2}$~mbar) and then refilled with argon. The reaction mixture was first heated up to 180$^{\circ}$C and then brought slowly up to 320$^{\circ}$C at a rate of 2$^{\circ}$C/min. The mixture was refluxed at 320$^{\circ}$C for 30~min. After the mixture was cooled down to room temperature, the NPs were precipitated with ethanol (or acetone) and collected by centrifugation. The NPs were washed by dissolving them in a non-polar solvent, such as hexane and reprecipitation with ethanol. This ``washing procedure" was repeated three times. The NPs were stored in hexane at 4$^{\circ}$C. With the control of the solvent, reaction time, temperature, and heating rate, spherical MnO NPs with an average size of 6~nm, 12~nm and 19~nm were synthesized, respectively. The as-prepared MnO NPs are covered with an oleic acid shell and have a size distribution of about 20~$\%$. 
\subsection{Synthesis of monomer FePt nanoparticles}\label{syn_FePt}
100~mg of Pt(acac)$_2$ and 800~\textmu L of oleic acid were dissolved under argon in 1-octadecene and degassed by the same procedure described for manganese oxide NPs. The reaction mixture was heated up to 120$^{\circ}$C, and 130~$\mu$L Fe(CO)$_5$ were injected. After 5 minutes, 800~$\mu$L of oleylamine were added, and the mixture was heated up to 150$^{\circ}$C and held at this temperature for 1~h. The NPs with an average size of 6~nm were washed and stored in hexane at 4$^{\circ}$C. In order to avoid the agglomeration, the FePt NPs are coated with an oleic acid shell.
\subsection{Synthesis of heterodimer FePt@MnO nanoparticles}\label{syn_dimer}
To prepare FePt@MnO heterodimer NPs, FePt NPs with the desired size were synthesized as described above and the MnO NPs were epitaxially grown on the surface of the FePt seed particles. As an example, 10~mg of as-prepared FePt NPs, 300~$\mu$L of oleic acid, 600~$\mu$L of oleylamine and 30-60~mg of manganese oleate were dissolved in 10~mL of hexadecane and degassed three times at 80$^{\circ}$C under reduced pressure (1$\times$10$^{-2}$~mbar) and then refilled with argon. The reaction mixture was heated up to 290$^{\circ}$C and then held there for 1~h. After cooling down to room temperature, NPs were washed by the procedure as described. The different amounts of manganese oleate regulate the size of the manganese oxide domain on the FePt surface. The as-prepared FePt@MnO heterodimer NPs with an average size of 6~nm@9~nm, 6~nm@11~nm, 6~nm@12~nm and 6~nm@16~nm, respectively, are covered with an oleic acid shell. The size distributions are slightly different for different samples. The FePt and MnO subunits have an average size distribution of 30~$\%$ and 25~$\%$, respectively.  Both FePt and MnO NPs are synthesized in spherical shape.

\begin{figure}
 \includegraphics[width=0.4\textwidth]{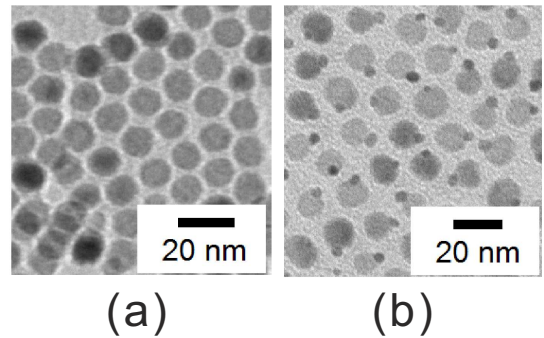}
 \caption{TEM image of (a) MnO NPs with 12~nm diameter, (b) FePt@MnO heterodimer NPs with 4~nm@11~nm size. \label{fig.NPs}}
 \end{figure}

The NPs were characterized by transmission electron microscopy (TEM). TEM images were recorded using a Philips EM420 microscope with an acceleration voltage of 120~kV. Samples for TEM were prepared by dropping a dilute solution of NPs in the appropriate solvent (hexane) onto a carbon coated copper grid (Plano, Wetzlar; Germany). Fig.~\ref{fig.NPs} shows TEM images of MnO NPs with an average size of 12~nm and the FePt@MnO heterodimer NPs with 4~nm@11~nm size. 

Magnetometry measurements were performed using a Superconducting Quantum Interference Device (SQUID) magnetometer (MPMS) and a Vibrating Sample Magnetometer (VSM) of the Physical Property Measurement System (PPMS) from Quantum Design. For magnetometry measurements, the NP dispersions were drop-casted onto phosphorus doped n-type $<$100$>$ silicon substrates at room temperature and dried for several hours. The sample sizes were 5~mm~$\times$~6~mm for PPMS measurements and 6~mm~$\times$~7~mm for MPMS measurements. The polarized neutron scattering was performed at the Diffuse Neutron Scattering (DNS) instrument \cite{Su2015,Schweika2001} using $\lambda$~=~4.2~\AA~at the Heinz Maier-Leibnitz Zentrum in Munich, Germany and the diffuse scattering spectrometer D7 \cite{Stewart2009} using $\lambda$~=~4.86~\AA~at the Institute Laue-Langevin in Grenoble, France. Both instruments are capable to use the XYZ-polarization analysis method to separate the nuclear coherent, nuclear spin incoherent, and magnetic contributions from the total measured scattering\cite{Scharpf1993,Schweika2010,Stewart2009}. For neutron scattering, 50~mg of dried MnO NPs powder are wrapped in aluminium foil and placed in a cylindrical aluminium sample holder. Single MnO NPs with 12~nm diameter were measured using the DNS-instrument. Four FePt@MnO dimer NP batches with about 10~mg each were measured at the D7 instrument. The FePt subunits had an average diameter of 6~nm, and the sizes of MnO subunits were 9~nm, 11~nm, 12~nm and 16~nm, respectively. The four FePt@MnO samples were dried separately on the aluminium foil. The aluminium foils were then folded like a ring of 2~cm diameter and 1.0~-~1.5~cm height. The samples were marked and placed into the sample holder at different heights to increase the footprint within the neutron beam.

\section{Results}\label{results}
\begin{figure}
	\includegraphics[width=0.4\textwidth]{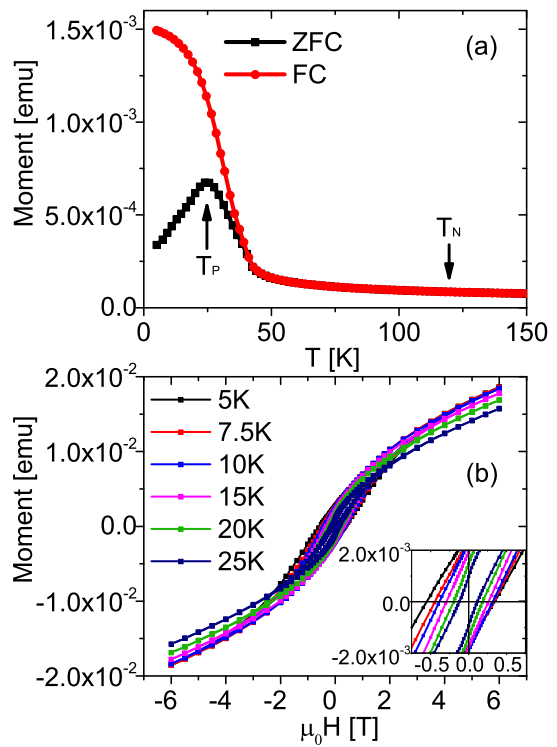}
	\caption{(a) ZFC/FC curves of 12~nm MnO NPs measured at 100~mT and (b) hysteresis loops of 12~nm MnO NPs measured at low temperatures. The inset shows a zoom-in around the origin.  \label{fig.zfc_ams354}}
\end{figure}

\begin{figure}
	\includegraphics[width=0.4\textwidth]{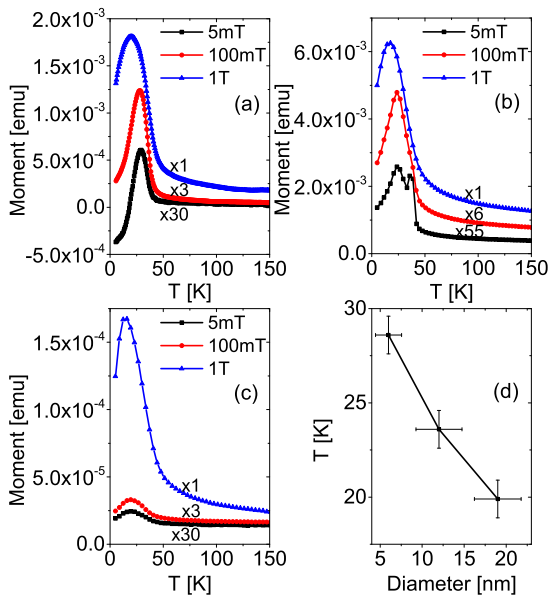}
	\caption{ZFC curves of (a) 6~nm, (b) 12~nm, (c) 19~nm MnO NPs. Panel (d) shows the peak temperatures in the ZFC curves as a function of the NP sizes measured at 100~mT. The negative values at low temperatures in (a) are possibly due to a negative residual magnetic field of the superconducting solenoid during the cooling procedure.  \label{fig.zfc_mno}}
\end{figure}

Temperature dependences of the magnetization were measured via the ZFC and FC procedure. After cooling the sample from high temperature above the N\'eel temperature of bulk MnO (118~K) to 5~K without an external magnetic field, the ZFC magnetization was measured during heating at various magnetic fields. The FC magnetization was achieved by measuring the magnetization while cooling the system in the presence of the same magnetic field. ZFC and FC magnetization curves for 12~nm MnO NPs are shown in Fig.~\ref{fig.zfc_ams354}(a). Around the bulk N\'eel temperature of MnO (118~K) both ZFC and FC curves show only a flat behavior both in the original data and in its derivative. Instead of the peak at the N\'eel temperature of bulk MnO, a peak appears at $T_P$~$\approx$~24~K in the ZFC magnetization of MnO NPs. This phenomenon in the ZFC magnetization is often observed in SPM systems \cite{Bedanta2013,Bedanta2015}. However, the field dependence of the peak temperature shown in Fig.~\ref{fig.zfc_mno} excludes this simple explanation. This will be discussed below in more depth.  The FC magnetization increases with a decrease in the temperature due to the freezing of the magnetic moments along the magnetic field. The ZFC and FC curves split around 40~K. Such splitting is often found in a SPM system or a spin glass due to the freezing of magnetic moments. In AF systems the splitting is also found due to the presence of pinned AF domain walls \cite{Montenegro1991}. 

In Fig.~\ref{fig.zfc_ams354}(b), hysteresis loops are plotted for MnO NPs at 5~K, 7.5~K, 10~K, 15~K, 20~K and 25~K. Each hysteresis loop was measured after the sample was cooled from high temperature above the N\'eel temperature of bulk MnO (118~K) at 100~mT magnetic field. The opening of the hysteresis loop may result from either spin canting effects, FM shell behavior \cite{Blundell}, blocked SPM \cite{Wiekhorst2003} or the presence of pinned AF domains \cite{Montenegro1991}. The centers of the hysteresis loops are shifted towards negative fields at low temperatures (Fig.~\ref{fig.zfc_ams354}(b) inset), which indicates an exchange bias (EB) effect. For pure AF ordering EB is not expected. However, on the surface of NPs a ``spin glass like" or ``FM-like" shell can possibly be observed due to the spin canting effect and spin disorder \cite{Bedanta2013,Winkler2008,Kodama1996}. The exchange coupling between the ``spin glass like" or ``FM-like" shell and the AF core could cause the EB effect in pure AF NPs \cite{Nogues2005}. However, the EB effect observed here is more likely to be caused by an oxidation of the NP surface from MnO to ferrimagnetic Mn$_3$O$_4$ or Mn$_2$O$_3$ \cite{Berkowitz2008,Golosovsky2009,Silva2013,Ragg2016}. Due to the exchange interaction between the AF MnO core and the ferrimagnetic shell, an EB effect can occur \cite{Nogues2005,Berkowitz2008,Golosovsky2009}. 

To further elucidate the origin of the ZFC peak, \textit{m} vs. \textit{T} measurements were performed under various applied fields. Fig.~\ref{fig.zfc_mno} displays the ZFC magnetic moment curves for various sizes of MnO NPs measured at magnetic fields of 5~mT, 100~mT and 1~T. Also here no feature is found near the N\'eel temperature of bulk MnO, but a peak at much lower temperature instead. The peak temperature shows a weak decrease with an increase of the magnetic field. Such weak field dependence is very different from the behavior found in SPM systems \cite{Yi2007,Kachkachi2000,Chantrell2000,Barbeta2010}. There, the blocking temperature decreases rapidly with the increase of the magnetic field in the order of few hundreds of mT. In contrast, the stability of the peak temperature against the magnetic field found here is only encountered in AF systems since the critical fields of most AFs are usually very high (tens to hundreds of Tesla) \cite{Benitez2008,He2007}. 

Moreover, the peak temperature in the ZFC magnetization shifts towards higher temperatures with decreasing NP size. This behavior is surprising because usually a decrease of transition temperatures is expected with decreasing length scales due to the finite size scaling effect \cite{He2007}. This also hints towards the interpretation that the ZFC peak does neither signify a phase transition nor a conventional SPM blocking temperature.  

In Fig.~\ref{fig.zfc_mno}(b), a second peak at ca.~40~K can be observed in the ZFC curve of 12~nm MnO NPs measured at 5~mT. This peak temperature matches the $T_C$ of Mn$_2$O$_3$ or Mn$_3$O$_4$, which reveals a possible stronger oxidation on the surface of these MnO NPs. 

In order to explain the unusual properties of MnO NPs observed in magnetometry, powder diffraction was performed using polarized neutron scattering at the DNS instrument. The intensity of the AF ($\frac{1}{2} \frac{1}{2} \frac{1}{2}$) Bragg peak and the nuclear (111) Bragg peak are measured above and below the ZFC peak temperature found from the magnetometry measurements and also above and below the bulk N\'eel temperature of MnO. The $<$111$>$ directions are along the AF propagation vectors. The nuclear (111) Bragg peak provides the structural information of MnO NPs, while the ($\frac{1}{2} \frac{1}{2} \frac{1}{2}$) Bragg peak reflects the AF magnetic structure of the system. The temperature dependence of the integrated intensity of the ($\frac{1}{2} \frac{1}{2} \frac{1}{2}$) Bragg peak is directly proportional to the square of the AF order parameter of the MnO NPs \cite{Feygenson2010}. 

With the help of polarization analysis, the magnetic scattering is hereby separated from the nuclear coherent and nuclear spin incoherent scattering \cite{Stewart2009,Schweika2010,Scharpf1993}. Fig. 4(a) shows the separated polarized neutron scattering intensity measured on 12~nm MnO NPs. The AF ($\frac{1}{2} \frac{1}{2} \frac{1}{2}$) Bragg peak of MnO is observed at $Q$~=~1.25~\AA$^{-1}$. The (111) nuclear Bragg peak of MnO is measured at $Q$~= 2.5 \AA$^{-1}$~as expected. Polarization analysis is necessary because the oleic acid shell covering the NPs contains hydrogen and produces significant nuclear spin incoherent scattering, which dominates the magnetic and the nuclear scattering. In particular for the extremely small amount of NP powder (i.e. 50~mg) only by this approach the measurements could be performed.

\begin{figure}
	\includegraphics[width=0.48\textwidth]{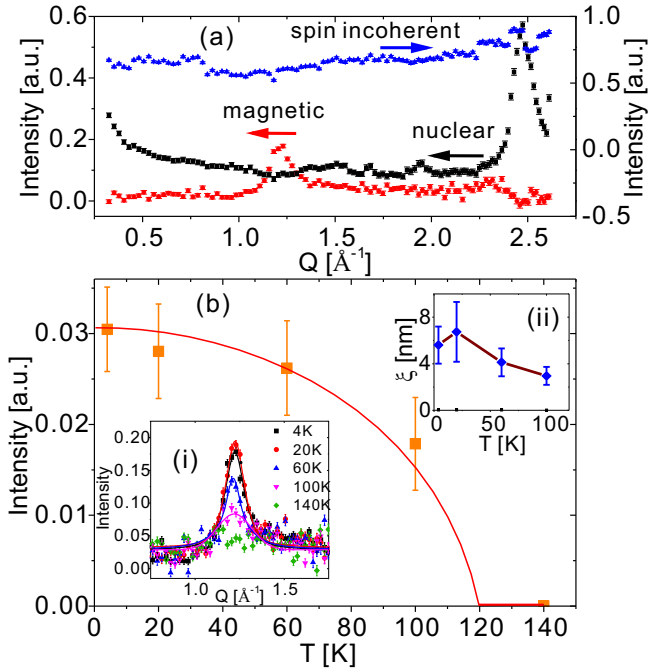}
	\caption{(a) Separated neutron scattering contributions of 12~nm MnO NPs from polarized neutron scattering measured at 4~K. The nuclear coherent (black squares), spin-incoherent (blue triangles) and magnetic (red circles) components of the MnO NPs measured at 4~K are shown. (b) Temperature dependence of the integrated intensity of the AF ($\frac{1}{2} \frac{1}{2} \frac{1}{2}$) Bragg peak. The orange squares represent the intensity of the AF ($\frac{1}{2} \frac{1}{2} \frac{1}{2}$) Bragg peak with error bars and the red line is a guide to the eye assuming the bulk $T_N$ of 120~K and a continuous transition\cite{Feygenson2010}. The inset (i) in (b) shows the magnetic ($\frac{1}{2} \frac{1}{2} \frac{1}{2}$) Bragg peaks at different temperatures. The inset (ii) in (b) displays the magnetic correlation length as function of temperature.  \label{fig.dns_mno}}
\end{figure}

Fig.~\ref{fig.dns_mno}(b) shows the integrated intensity of the AF ($\frac{1}{2} \frac{1}{2} \frac{1}{2}$) Bragg peak for 12~nm MnO NPs measured at various temperatures. The intensity of the peak decreases monotonically with increasing temperature until it vanishes between 100~K and 140~K, i.e. near the bulk N\'eel temperature of MnO at 118~K. More data points and thus a more precise determination of the transition temperature were not possible due to the extremely small amount of particles and thus extensive integration times. The intensity of the nuclear (111) peak vs. temperature shows the expected constant behavior inside the error bars (data not shown).

The magnetic Bragg peaks of the MnO NPs were fitted using a pseudo Voigt function. The DNS- instrument resolution is considered as width of the Gaussian profile, and the broadening of the Lorentz profile is due to the NPs. One hereby also obtains the magnetic correlation lengths $\xi$ of the MnO NPs using the Scherrer formula \cite{Scherrer1918}. The temperature dependence of the magnetic correlation length is shown in the inset (ii) of Fig.~\ref{fig.dns_mno}(b). The magnetic correlation length of 12~nm MnO NPs is 6-7~nm at low temperature. It decreases with increasing temperature very likely due to thermal fluctuations. This reduced value for the magnetic correlation length compared to the diameter of the NPs is either due to the presence of a Mn$_2$O$_3$ or Mn$_3$O$_4$ shell,  or a non-AF ordered MnO shell (e.g. due to canting or frustration) or due to an AF domain state. The grain size of the NPs calculated from Scherrer formula is about 13~nm from the nuclear peak, which matches the size of the NPs.

\begin{figure}
	\includegraphics[width=0.48\textwidth]{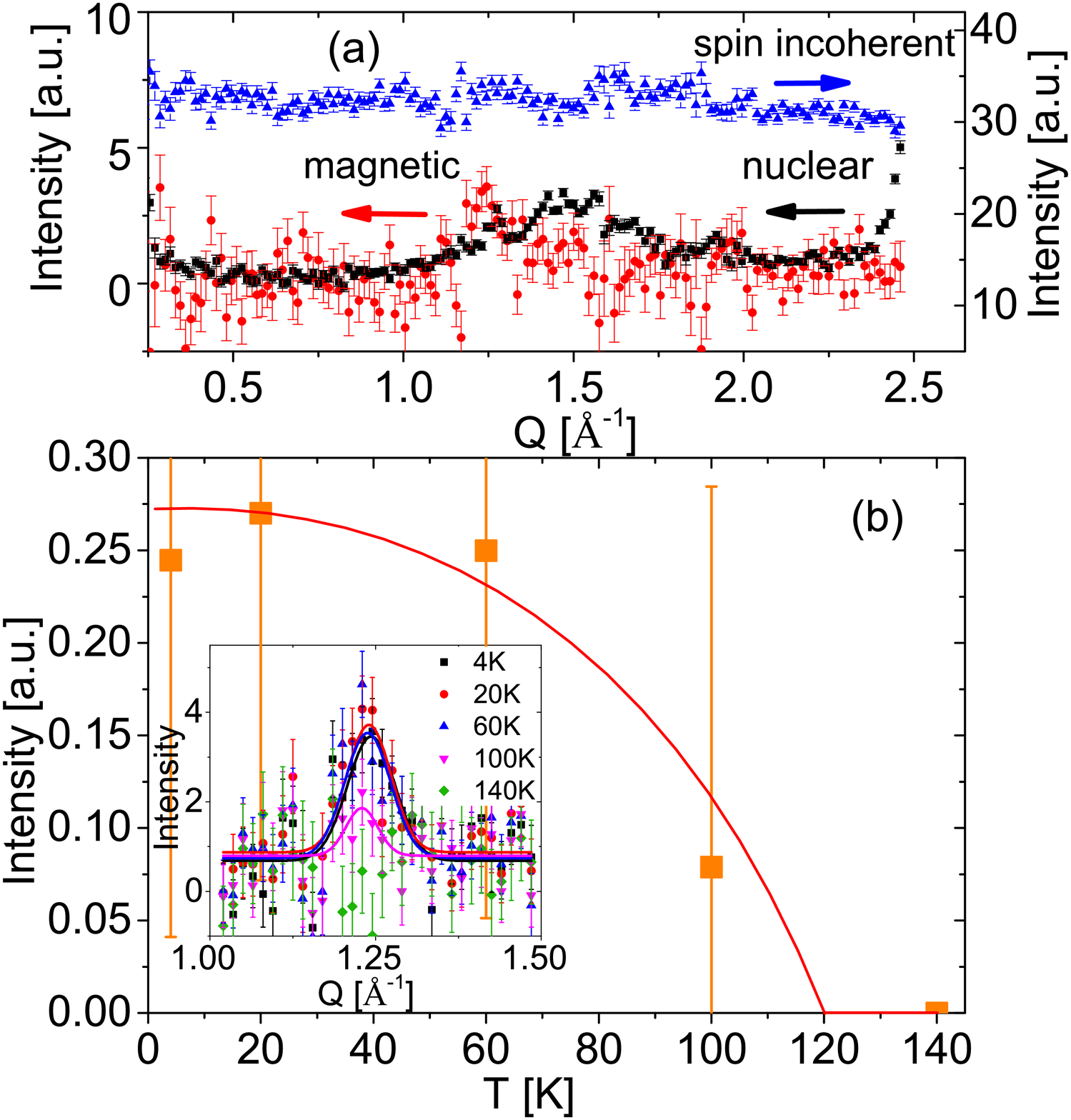}
	\caption{(a) Separated contributions from polarized neutron scattering measurements on FePt@MnO NPs at 4~K. (b) Temperature dependence of the integrated intensity of the AF ($\frac{1}{2} \frac{1}{2} \frac{1}{2}$) Bragg peak of the MnO subunits. The squares represent the intensity of the AF ($\frac{1}{2} \frac{1}{2} \frac{1}{2}$) Bragg peak with error bars and the red line is a guide to the eye assuming the bulk value of $T_N$. The inset in (b) shows AF ($\frac{1}{2} \frac{1}{2} \frac{1}{2}$) Bragg peak at various temperatures.  \label{fig.dns_fmo}}
\end{figure}

To understand the magnetic behavior of MnO NPs it is interesting to study the magnetic influence of a FM particle in direct contact to the MnO NP. To this end FePt@MnO dimer NP samples are measured. However, the mass of individual samples was insufficient for a meaningful neutron scattering experiment. Therefore, the four samples with NP sizes (1) 6~nm@9~nm, (2) 6~nm@11~nm, (3) 6~nm@12~nm and (4) 6~nm@16~nm  were combined. The samples were dried separately on aluminium foil, wrapped and marked independently and placed into the sample holder at different heights to increase the footprint within the beam. The measured data thus represented an ensemble average of the signal from the four samples. The neutron scattering measurements on the FePt@MnO dimer NPs were performed at the same temperature values as for the MnO NPs. Fig.~\ref{fig.dns_fmo}(a) shows the separated neutron scattering contributions measured at D7 instrument (ILL, Grenoble). The error bars are relatively large because of the extremely small amount of sample (ca. 40~mg) and the relative broad size distribution of the NPs. 

At $Q$~=~1.25~\AA$^{-1}$, a weak magnetic ($\frac{1}{2} \frac{1}{2} \frac{1}{2}$) Bragg peak can be observed. This magnetic peak matches the AF Bragg peak of the MnO NPs. The nuclear (111) peak of MnO is expected at $Q$~=~2.5~\AA$^{-1}$. Because of the limited $Q$ range, only an increase at $Q$~=~2.5~\AA$^{-1}$ is observed in the nuclear scattering of FePt@MnO NPs. Fig.~\ref{fig.dns_fmo}(b) shows the temperature dependence of the intensity of the AF Bragg peak of FePt@MnO NPs. The magnetic peaks of the FePt@MnO NPs were fitted with Gaussian function. The intensity shows an AF order parameter behavior and vanishes at a temperature between 100~K and 140~K. This result is similar to the order parameter curve measured for single MnO NPs. The FePt subunits in the FePt@MnO heterodimers show no obvious influence onto the AF order parameter of MnO NPs.

\begin{figure}
	\centering
	\subfloat{\includegraphics[width=0.43\textwidth]{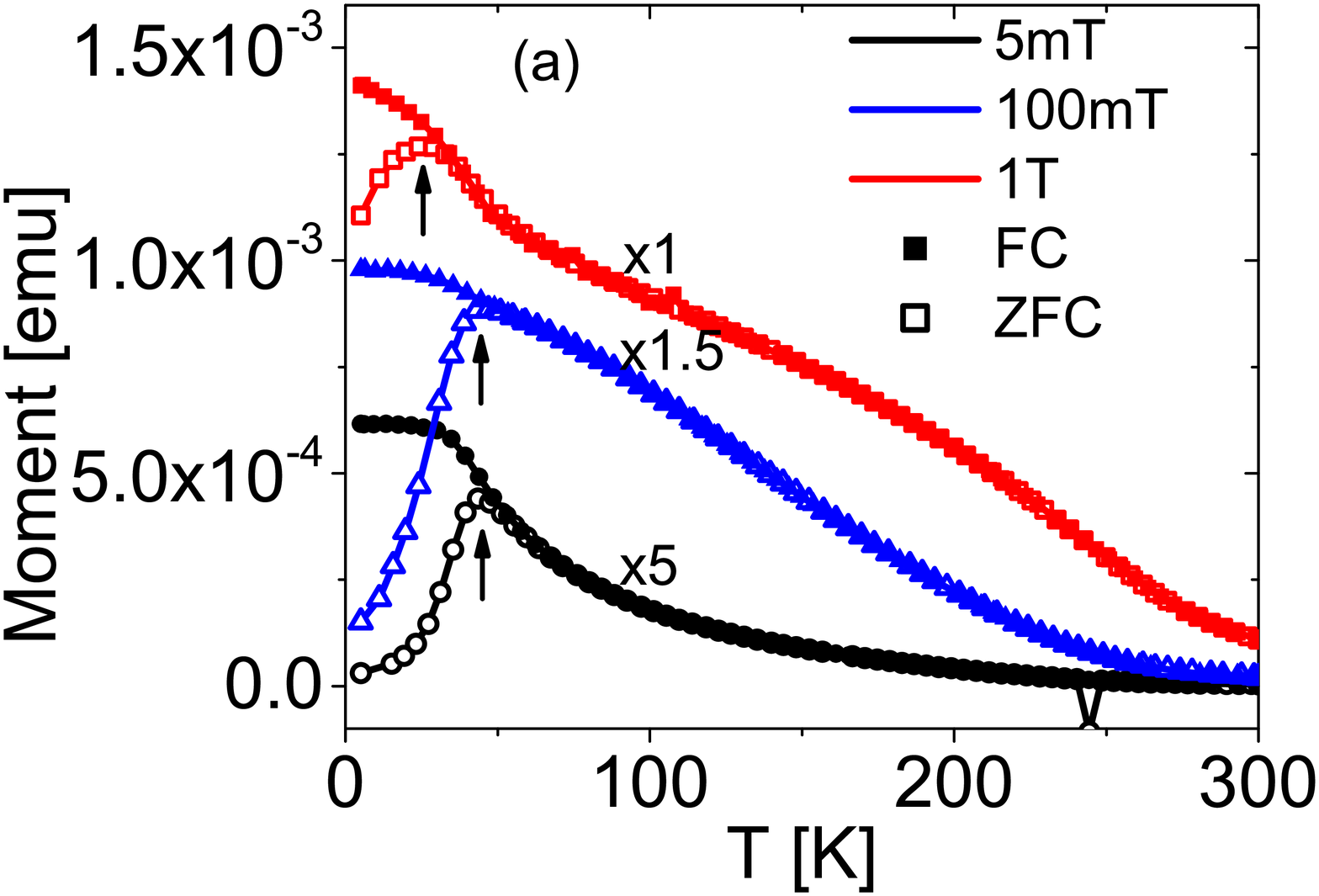}}\\ \subfloat{\includegraphics[width=0.435\textwidth]{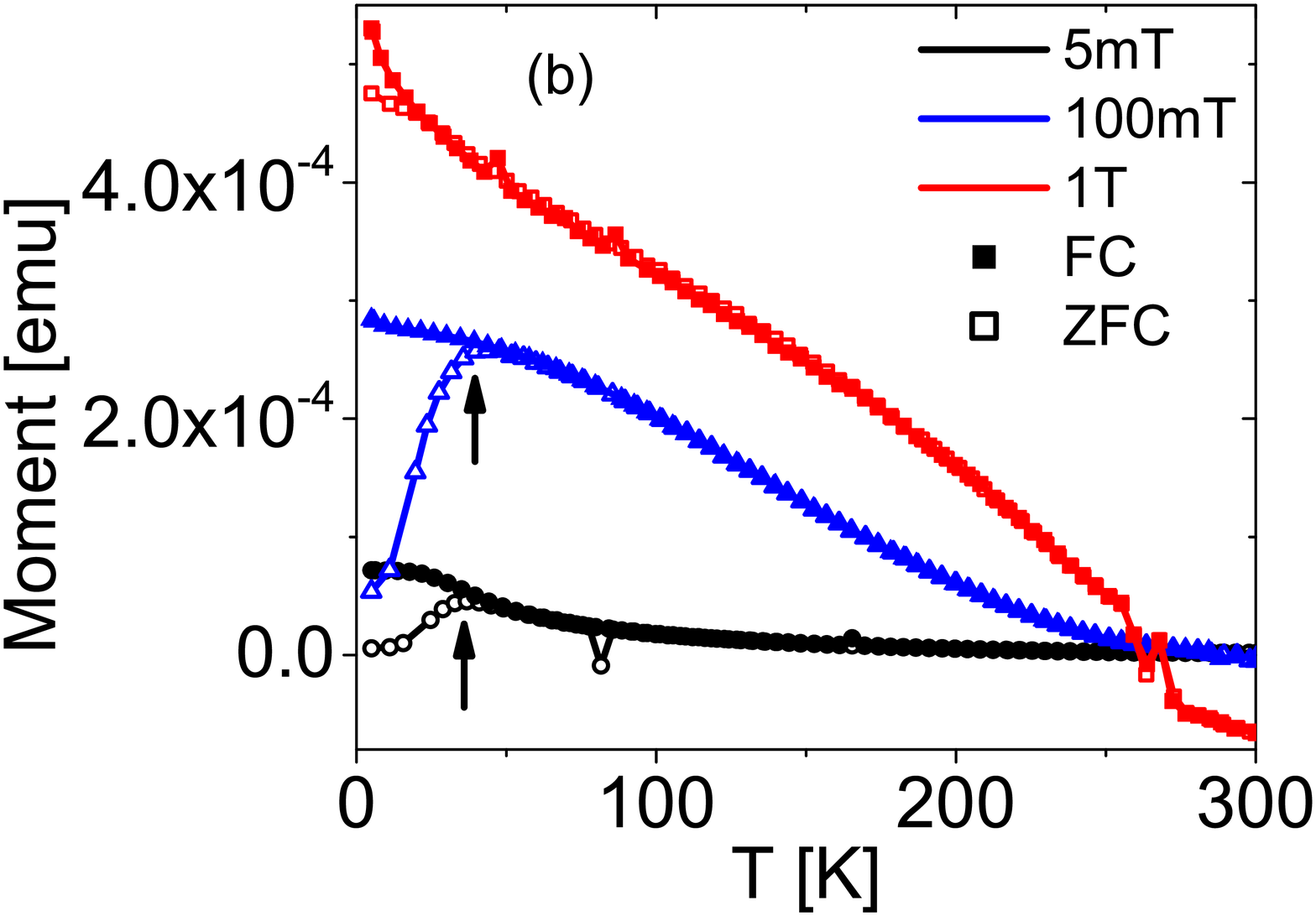}}
	\caption{ZFC (open symbols) and FC (solid symbols) curves of FePt@MnO NPs with (a) 6~nm@12~nm and (b) 6~nm@9~nm size measured at different magnetic fields, H~=~5~mT (black), 100~mT (blue), and 1~T (red), respectively.}\label{fig:zfc_fmo}
\end{figure}

The influence of the FePt subunits onto the AF order of MnO NPs is also studied using magnetometry. In Fig.~\ref{fig:zfc_fmo}, ZFC/FC magnetization curves of FePt@MnO NPs with (a) 6~nm@12~nm and (b) 6~nm@9~nm sizes are shown. As for single MnO NPs, the N\'eel temperature of MnO at 118~K cannot be observed in either ZFC or FC curves of all FePt@MnO NP systems. Instead, a peak appears at 44~K, 46~K and 26~K in the ZFC curve for the 5~mT, 100~mT and 1~T measurements, respectively for 6~nm@12~nm FePt@MnO NPs. Moreover, as the size of MnO subunit decreases, exchange bias effect to stabilize the magnetic moments in FePt NPs reduces. Therefore, the peak disappear at 1~T for 6~nm@9~nm FePt@MnO NPs. Since there are not two separate peaks found in the ZFC curves, the FePt subunits and the MnO subunits are strongly coupled. While for the MnO monomers, only a very weak temperature dependence is found above 50~K, the dimers show a pronounced increase with decreasing temperature starting already at room temperature. We attribute this behavior to the FePt subunit. Compared to single MnO with 12~nm size, the peak temperature increases (e.g. from 24~K to 46~K at 100~mT) due to the exchange biased FePt NPs. For small magnetic field, the magnetic moment of FePt is much larger than that of MnO, so that the result is dominated by the FePt subunits. The $T_P$ of FePt decreases rapidly with increasing magnetic field. Hence, at high fields the peak is very likely only due to the MnO subunits.

\begin{figure}
	\includegraphics[width=0.4\textwidth]{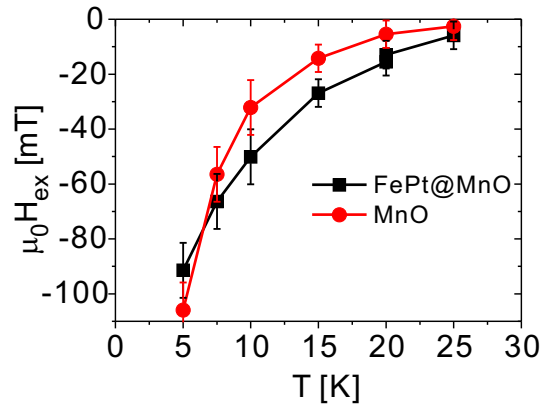}
	\caption{Exchange bias field vs. \textit{T} obtained from hysteresis loops at various temperatures after field cooling in 100~mT measured on 12~nm MnO NPs (red circles) and 6~nm@12~nm FePt@MnO NPs (black squares). The lines are guide to the eye. \label{fig.Hex}}
\end{figure}

On both systems, i.e. on MnO monomer and also on FePt@MnO dimer NPs, an exchange bias effect has been observed. The exchange bias fields $H_{ex}$ as function of temperature are plotted in Fig.~\ref{fig.Hex}. $|H_{ex}|$ drops quickly with increasing temperature for both MnO and FePt@MnO NPs, and reaches zero at approximately 25~-~30~K. For monomer MnO NPs $|H_{ex}|$ decreases faster and vanishes at a slightly lower temperature ($\approx$25~K) than those exchange biased by FePt NPs ($\approx$30~K). Interestingly, these temperatures match approximately the ZFC peak temperatures measured in the ZFC magnetization curves until 1~T inferring that the ZFC peak and the breakdown of exchange bias are indeed correlated.

\section{Discussion}\label{discussion}
Summarizing our findings on the MnO NPs so far, we arrive at a puzzling scenario. I.e. in magnetometry a peak at low temperatures ($\approx$~25~K) is observed in the ZFC magnetization curve of MnO NPs instead of a feature marking the N\'eel temperature at 118~K. However, in the polarized neutron scattering measurements, the expected AF order parameter behavior of MnO is confirmed with the regular N\'eel temperature near 118~K. Consequently, the ZFC peak at $\approx$~25~K cannot mark a finite size scaled phase transition into the AF state of MnO. 

Moreover, the ZFC peak does not signify simple SPM behavior, since such an explanation contradicts to the field dependence of the ZFC curves \cite{Yi2007,Kachkachi2000,Chantrell2000,Barbeta2010}. I.e. the peak temperature shifts very weakly even up to large applied fields of 1~T. Thus the ZFC peak must have another origin.
To understand this behavior, we have to consider the two different time scales, at which the measurements are performed both during magnetometry and during neutron scattering. While magnetometry probes the magnetic state on a typical time scale of 10~s (integration time per point), neutron scattering has a probing time scale in the order of ps \cite{Angst2012}. In addition, while the magnetic moment from magnetometry yields information about the quasistatic net magnetization configuration, the polarized neutron scattering data provide direct information about the spin ordering (e.g. AF ordering in case of the ($\frac{1}{2} \frac{1}{2} \frac{1}{2}$) Bragg peak).

\begin{figure}
	\includegraphics[width=0.4\textwidth]{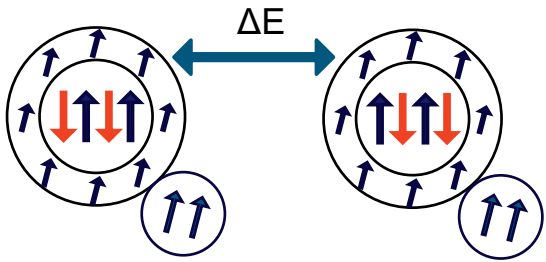}
	\caption{Model to explain the magnetism in FePt@MnO NPs: the MnO core shows antiferro-SPM (AF-SPM) behavior while it is exchange coupled to a ferrimagnetic Mn$_2$O$_3$ or Mn$_3$O$_4$ shell. This core-shell system is exchange coupled to a FM FePt NP (smaller circles). \label{fig.dimer_model}}
\end{figure}

A possible conclusion is that the MnO NPs exhibit three different regions of magnetic behavior depending on the temperature: \\
1. Above $T_N \approx$~118~K the system is paramagnetic. The \textit{m} vs. \textit{T} curves show a Curie-Weiss type behavior, while neutron scattering displays zero intensity in the AF Bragg peak. Therefore both measurement techniques are consistent.\\
2. For $T_P < T < T_N$ magnetometry probes a fluctuating system macroscopically similar to the unblocked SPM state \cite{Petracic2010}. However, neutron scattering evidences clearly AF short range order inside the MnO NPs. One explanation of this finding is that the system is in a SPM-type of state where the N\'eel-vector of AF ordering fluctuates thermally induced. Such a state could be called antiferro-SPM (AF-SPM) state (Fig.~\ref{fig.dimer_model}). This also explains why no feature at the N\'eel temperature is observed in magnetometry. The results from the two methods are then also consistent since the magnetometry probes at several orders of magnitude larger time scales ($\sim$s) than neutron scattering ($\sim$ps). While magnetometry `sees' only fluctuating AF-SPM behavior, the neutron scattering `sees' AF ordering. It would be interesting to observe exactly such an onset of ``dynamical ordering" by precise temperature steps and much better statistics. However, such an experiment is challenging due to the extremely small sample mass.\\
3. At $T < T_P$ both magnetometry and neutron scattering observe a ``blocked" AF state. In magnetometry the crossover from the second to third regime is marked by a peak similar to a SPM \cite{Petracic2010}. In neutron scattering no change occurs apart from the monotonous increase corresponding to the AF order parameter.

Consequently the AF-SPM model seems to be the model of choice. Also the field dependence of the ZFC peak position can be explained by this model, since the underlying AF state would also show a weak field dependence as a regular AF.

In this case, the peak temperature is estimated according to a simple N\'eel-Brown ansatz:
\begin{equation*}
\tau_{exp} =\tau_0 \; exp \left( \frac{\Delta E}{k_B \; T_P} \right) \\
\leftrightarrow T_P = \frac{\Delta E}{k_B \; ln(\tau_{exp}/\tau_0)}
\end{equation*}
with an energy barrier $\Delta E = K \times N$, \textit{K} the single ion anisotropy of Mn ions, $N$ the number of Mn ions inside a NP, $N \approx$ 1000, $\tau_0 = 10^{-10}~s$ and $\tau_{exp} = 10~s$. However, the observation from Fig.~\ref{fig.zfc_mno}(d) contradicts this explanation. I.e. the peak temperature shifts to smaller values with increasing particle size. From the above model one would rather expect a positive proportionality of size and $T_P$. One possible explanation would be that larger NPs can contain more independently fluctuating AF domains and thus a larger particle could exhibit an effectively smaller energy barrier for AF flipping than a smaller particle. 

One should note that similar studies on MnO NPs reveal the importance of an oxidized shell \cite{Berkowitz2008,Golosovsky2009}. One criterion is the color of the suspension. A greenish suspension signifies pure MnO NPs, whereas a brownish one hints to the presence of a ferrimagnetic Mn$_3$O$_4$ or Mn$_2$O$_3$ shell \cite{,Schladt2009,Kim2005}.

Our dispersion has a brownish color. Hence a Mn$_3$O$_4$ or Mn$_2$O$_3$ shell must be present. The presence of a ferrimagnetic shell (Fig.~\ref{fig.dimer_model}) can explain the observed exchange bias effect \cite{Nogues2005}, which cannot be understood using the AF-SPM model alone.

Therefore we conclude that the magnetic behavior of MnO NPs needs to be explained by both models, i.e. AF-SPM \textit{and} a core-shell structure. The heterodimer system FePt@MnO shows a qualitatively similar behavior. While the FePt subunit is governed by SPM behavior \cite{Sun2006,Sun2000}, the MnO unit shows the behavior mentioned above. Moreover, an additional exchange bias effect occurs due to exchange interactions between the FePt particle and the ferrimagnetic shell of the MnO NP. This yields a slight increase in the EB value as shown in Fig.~\ref{fig.Hex}.

\section{Conclusion}\label{conclusion}
The AF order parameter measured by polarized neutron scattering shows a phase transition temperature around 118~K near to the N\'eel temperature of bulk MnO. This result confirms the existence of AF ordered MnO. However, magnetometry reveals a peak in the ZFC curves at lower temperatures (ca.~25~K) and shows no sign of ordering at $T_N$. This seemingly contradicting behavior can be explained by two effects occurring simultaneously in MnO NPs. First, there is AF-SPM behavior, where the single NP exhibits short range AF order below $T_N$, but thermal fluctuations destroy the macroscopic magnetometric signal above $T_P \approx$~25~K. Below $T_P$ the system enters a blocked AF state, which is signified by a peak in the ZFC curve. Second, the MnO NPs have a ferrimagnetic shell so that EB is observed between the AF core and the shell. The behavior of the heterodimer system is governed by the superposition of SPM behavior of the FePt unit, the AF-SPM + core-shell behavior of the MnO and an exchange bias between FePt and the MnO-shell.

This study provides insight into the magnetic structure of MnO NPs as well as the magnetic behavior of an FePt@MnO heterodimer NP systems. It also stresses the importance to characterize the magnetic behavior of magnetic NPs in detail, if they are to serve as building blocks for novel multifunctional materials.  It will be interesting to compare Monte Carlo simulations of both single MnO NPs and FePt@MnO heterodimer NPs with experimental results. 

\begin{acknowledgments}
We thank W. Schweika and J. Voigt for fruitful discussions. We thank Berthold Schmitz for technical support. HB, OK, AS and WT thank the SFB 1066 ``Nanodimensionale polymere Therapeutika f\"ur die Tumortherapie" for partial support.
\end{acknowledgments}

\bibliography{PRB}

\end{document}